\documentclass[preprint,12pt,3p]{elsarticle}
\makeatletter
\def\ps@pprintTitle{%
 \let\@oddhead\@empty
 \let\@evenhead\@empty
 \def\@oddfoot{}%
 \let\@evenfoot\@oddfoot}
\makeatother

\usepackage{amssymb}
\usepackage{esint}
\usepackage{algorithm}
\usepackage{algorithmic}
\usepackage[]{hyperref}

\begin{document}

\begin{frontmatter}

\title{Linear layout of multiple flow-direction networks for landscape-evolution simulations} 

\address[label1]{Department of Civil and Environmental Engineering, Princeton University, New Jersey, NJ 08544, USA}
\address[label2]{ Princeton Environmental Institute, Princeton University, Princeton, New Jersey 08544, USA}
\address[label3]{Princeton Institute for International and Regional Studies, Princeton University, New Jersey 08544, USA}

\cortext[cor1]{I am corresponding author}

\author[label1]{Shashank Kumar Anand}
\ead{skanand@princeton.edu}
\author[label1,label2,label3]{Milad Hooshyar}
\ead{hooshyar@princeton.edu}
\author[label1,label2]{Amilcare Porporato \corref{cor1}}
\ead{aporpora@princeton.edu}

\begin{abstract}
We present an algorithm that is well suited to find the linear layout of the multiple flow-direction network (directed acyclic graph) for an efficient implicit computation of the erosion term in landscape evolution models. The time complexity of the algorithm varies linearly with the number of nodes in the domain, making it very efficient. The resulting numerical scheme allows us to achieve accurate steady-state solutions in conditions of high erosion rates leading to heavily dissected landscapes. We also establish that contrary to single flow-direction methods such as D8, D$\infty$ multiple flow-direction method follows the theoretical prediction of the linear stability analysis and correctly captures the transition from smooth to the channelized regimes. We finally show that the obtained numerical solutions follow the theoretical temporal variation of mean elevation.
\end{abstract}


\end{frontmatter}


\section{Introduction}
\label{S1}
The spatial organization of ridges and valleys on earth and other planets serve as the footprint of various processes such as weathering, erosion, sedimentation, creep, and tectonic uplift, etc. \cite{carr2000meter, chen2014landscape, fowler2011mathematical,  seybold2018branching, tomasko2005rain}. Their relative value controls the landscape profile from smooth to heavily dissected one with complex topographies. Landscape evolution models (LEMs) have been developed to explain the role of these processes on the formation and evolution of the Earth's surface \cite{birnir2001scaling, coulthard2001landscape, tucker2001channel, istanbulluoglu2005vegetation, koons1989topographic, perron2012root, roering2008well, smith1972stability, willgoose1991coupled}. LEMs aim to simulate the dynamics of land surface over large spatial and temporal scales (e.g., in the order of square-km and million years), where solving mass and momentum equations of water flow over the surface becomes computationally impractical. At such large spatiotemporal scales, the assumption of uniform precipitation over the domain and constant water velocity at every point in the direction of steepest descent are suitable, leading to minimalist models of landscape evolution \cite{chen2014landscape, fowler2011mathematical}. As a result, the water elevation term in LEMs is replaced by the drainage area, thereby bypassing the need for employing water transport equations explicitly.

Various flow-direction methods have been developed to compute the drainage area and produce the flow network in a computationally inexpensive way \cite{qin2007adaptive, costa1994digital, freeman1991calculating, holmgren1994multiple, o1984extraction, paik2012simulation, quinn1991prediction, tarboton1997new,bonetti2020channelization}. The choice of flow-direction method affects the type of flow network and calculation of the drainage area, which in turn affects the erosion term in the LEM and the accuracy of the solution. Implicit algorithms have been applied to solve the stream-power equation (the erosion term) mostly for the single flow-direction method \cite{barnes2019accelerating, braun2013very}. These algorithms have allowed obtaining efficient simulations of landscape evolution.

In this work we build upon these contributions, extending them to include multiple flow-direction methods, such as D$\infty$, which provide a better approximation of the specific drainage area (as the surrogate of water flux at a point, see Section \ref{S2}), compared to single flow-direction methods, such as D8 \cite{ gallant2011differential, goodchild1996gis, pan2004comparison, wolock1995comparison}. Contrarily to single flow-direction methods, multiple flow-direction methods generate a tangled flow network across the domain, which makes it difficult to find the linear ordering of nodes in the network necessary for efficient computation in the implicit solver. For this reason, LEMs using a multiple flow-direction method have been solved explicitly in time \cite{hooshyar2019interbasin, perron2008controls, roering2008well}. The explicit solver however poses a strict constraint on the time-step and makes it computationally impractical to solve LEM for highly dissected landscapes.

The paper is organized as follows. We present the governing equations for the LEM in the detachment-limited condition in Section \ref{S2}. In Section \ref{S3}, we show the difference between network traversal in single and multiple flow-direction networks for an efficient implicit computation of the erosion term. In Section \ref{S4}, we describe the proposed algorithm with a worked-out example using D$\infty$ method. In Section \ref{S5}, we provide the results of various numerical experiments performed using the proposed algorithm and evaluate the accuracy of obtained results by comparing them with theoretical solutions. We show the simulation results obtained using M8 method to indicate the scope of presented algorithm in \ref{appendix-sec1}. We finally present the pseudocode for the D$\infty$ method to show the implementation of the algorithm in \ref{appendix-sec2}.

\section{Governing equations}
\label{S2}

We focus on the detachment-limited conditions which assume the resistance to incision is the limiting condition for erosion rate rather than the hauling capacity of the channel to carry the material out of the domain so that the eroded material does not get redeposited within the domain \cite{howard1994detachment, izumi1995inception}. Under these premises, the temporal dynamics of land surface elevation $z$  is described as
\begin{equation}
\label{lem_eq}
\frac{\partial z}{\partial t}=D\nabla^2 z-K a^m |\nabla z|^n + U,
\end{equation}
where $D$ is the creep-diffusion coefficient, $K$ is an erosion coefficient, $m$ and $n$ are model parameters \cite{whipple1999dynamics}, $U$ is the uplift rate and $a$ is the specific drainage area. The parameters $D$, $K$, $m$, $n$ and $U$ are assumed constant, while the equation for the variable $a$ will be given in what follows.

The first term on the right-hand side (RHS) of equation (\ref{lem_eq}) is the sediment flux due to soil creep. The assumption of soil-creep flux being proportional to the topographic gradient recasts this term as a linear diffusion term \cite{ culling1960analytical, culling1963soil}. The second term on the RHS of equation (\ref{lem_eq}) represents the sediment flux due to water erosion. In detachment-limited condition, it is usually supposed to be proportional to the energy expenditure rate of the stream, giving it the form of a nonlinear, nonlocal sink term \cite{kirkby1971hillslope, seidl1992problem, seidl1994longitudinal, perron2008controls}. The nonlocality in erosion term is due to the presence of the specific drainage area, which makes the boundary conditions crucial in this model. Tectonic uplift is the external forcing acting beneath the surface, which is modeled as a constant source term on the RHS of equation (\ref{lem_eq}).

\begin{figure}[!hbt]
\centering
\includegraphics[width = \linewidth]{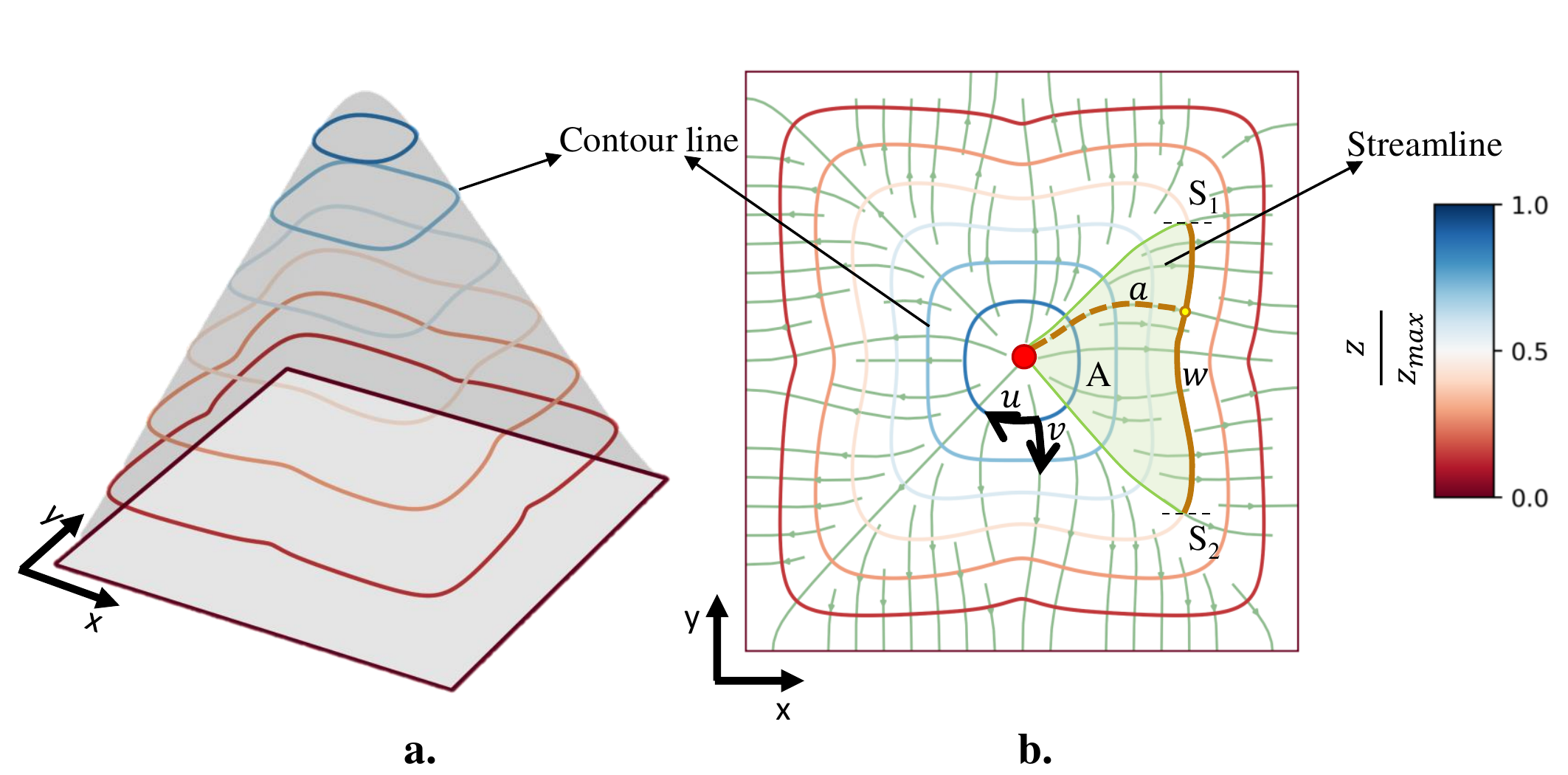}
\caption{Definition of the main variables. (a) Landscape surface near the first channel instability with six contour lines over a square domain (red delineating points at low altitude and blue depicting points at the high altitude). (b) The horizontal projection of the landscape with streamlines (in green) perpendicular to the projected contour lines. The schematic definitions of $a$ and $A$ are shown for a contour segment S1 --- S2. Starting from the hilltop (red point), the light green area between two streamlines manifests $A$ for the contour segment. Dotted brown streamline represents $a$ for a point on the contour segment. ($x,y$) is the Cartesian coordinate system and ($u,v$) is the curvilinear coordinate system where $u$ points along the contour lines of elevation field and $v$ points along the streamlines (gradient of the elevation field).}
\label{fig:one}
\end{figure}

As shown in Figure \ref{fig:one}, the total drainage area ($A$) is the horizontal projection of the area that flows to a finite portion of the contour line. Correspondingly, the specific drainage area ($a$) for a point in the domain is defined for a point as $\lim_{w\to 0} A/w$ where $w$ is the length of contour line passing through that point \cite{bonetti2018theory, gallant2011differential} and $A$ is the corresponding contributing area (see Figure \ref{fig:one}). As shown in \cite{bonetti2018theory}, the specific drainage-area equation is
\begin{equation}
\label{a_eq}
-\nabla \cdot \left(a\frac{\nabla z}{|\nabla z|}\right) = 1.
\end{equation}
Since the negative gradient vector points toward the steepest descent, the flow-velocity direction is parallel (opposite in sign) to the gradient vector as written in equation (\ref{a_eq}). This is apparent from the erosion term in equation (\ref{lem_eq}), where the slope of flow direction at a point in the domain is the magnitude of the (negative) gradient vector of the elevation field.

For a domain with typical length $l$, the two coupled-PDEs are non-dimensionalized in \cite{bonetti2020channelization} to obtain a dimensionless quantity 
\begin{equation}
\label{chi_eq}
\mathcal{C_I} = \frac{Kl^{m+n}}{D^nU^{1-n}},
\end{equation}
whose value indicates the tendency to form channels (low value indicates a smooth profile, while high value implies a dissected landscape; for this reason, we refer to it as the `channelization index').

Equations (\ref{lem_eq}) and (\ref{a_eq})
form a closed system of nonlinear partial differential equations to be solved with suitable initial and boundary conditions. These equations can be solved analytically in a special case ($m=n=1$) for non-dissected geometries in simple domains (see \cite{bonetti2020channelization} and Section \ref{SS511}). For channelized cases in complex topographies, the governing equations must be solved numerically. From a numerical point of view, the existence of a flow network draining the entire landscape presents a way to employ it for the drainage-area computation, which is needed every time the landscape elevation surface is updated. In the absence of this crucial information, the direct numerical solution of equations (\ref{lem_eq}) and (\ref{a_eq}) would be much more complicated. Thus, different flow-direction methods (D8, M8, D$\infty$, etc.) have been used to provide a numerical approximation of the underlying flow network and compute the drainage area ($A$). Assuming grid size to be adequate to give a good approximation of flow width, the field of $a$ is then approximated as $A/\Delta x$, where $\Delta x$ is the grid size \cite{chirico2005definition, tarboton1997new}, thus indirectly solving equation (\ref{a_eq}) numerically over the discretized domain. The flow network obtained at each time step is then used in the discretized form of equation (\ref{lem_eq}) to update the elevation at each point in the domain.

\section{Flow-direction methods and an efficient implicit calculation}
\label{S3}

The advantage of having no restrictions on the size of time-step in implicit formulations is counteracted by an expensive computation of the coupled nonlinear equations at every time-step \cite{langtangen2017finite}. We address this issue by decoupling the equations by transforming the system into upper/lower triangular.
We follow here the approach by \cite{barnes2019accelerating} and \cite{braun2013very} for single flow-direction method, which we extend to multiple flow-direction method by traversing the network (linear layout) in a way that the number of operations to update elevations at a time-step varies linearly with the total number of nodes. The algorithm to construct this linear layout of the flow network for an efficient implicit computation of the erosion term depends on the connectivity among the nodes, which in turn depends on the type of flow-direction method.

\begin{figure}[!hbt]
\centering
\includegraphics[width = \linewidth]{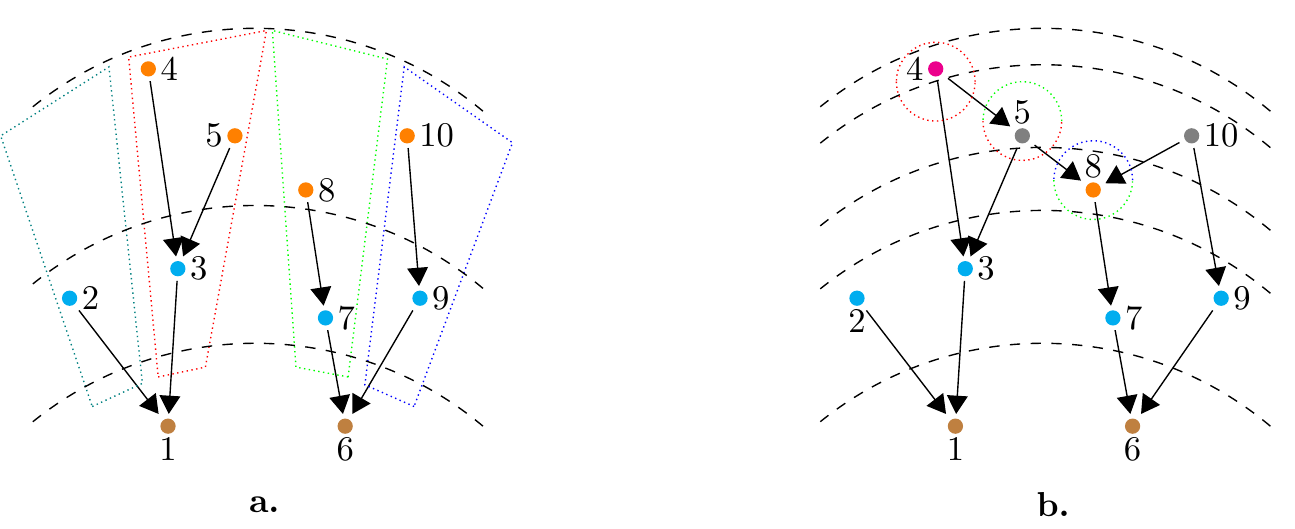}
\caption{Types of flow network. (a) Single flow-direction network with in-trees having nodes $1$ and $6$ as the respective sink nodes. Same-colored nodes belong to the same level with dotted black lines partitioning different levels. Four discrete colored regions show the independent flow lines. (b) Multiple flow-direction network with directed acyclic graphs having nodes $1$ and $6$ as the respective sink nodes and same-colored nodes indicating nodes at the same level. Three dotted circles indicate the intermixing of the flow lines that were not communicating with each other in panel (a). Modified from \cite{barnes2019accelerating}. Table \ref{table1} contains the relevant data structures used in the algorithm for multiple flow-direction network shown in panel (b).}
\label{fig:two}
\end{figure}

In a single flow-direction method, the flow from a node can only go to one node downstream of it \cite{koons1989topographic}. A node in the flow network, therefore, can have multiple donors (upstream nodes draining to the node) but must have a single receiver (node to which this node is passing its flow). This means that there exists a one-to-many relationship among nodes in the flow network looking from downstream to upstream with a unique path from each node to a node without any receiver (sink node). This layout is an anti-arborescence or in-tree tree which is a directed tree with root $r$ such that there exists a unique path directed from any node $n$ to the root $r$ \cite{hoque2016trees}. Figure \ref{fig:two}(a) shows the typical formation of anti-arborescence trees with multiple outlets acting as the sink nodes of the respective trees. The level of a node is defined as the number of direct edges between that node and the sink. Sink node forms the first level of the tree, sink's donors are on the second level, donors of sink's donors occupy the third level, and so on. In \cite{barnes2019accelerating,braun2013very}, different linear layouts of the in-tree structure (depth-first and breadth-first traversal respectively) have been employed to propose efficient implicit algorithms for a single flow-direction network.

For a multiple flow-direction method, a node can have multiple donors as well as multiple receivers, as shown in Figure \ref{fig:two} (b). Several flow lines can intersect and diverge in this paradigm, which leads to the intermingling of various branches of the in-tree structure \cite{tarboton1997new}. This framework does not remain an in-tree any more as it displays a directed acyclic graph \cite{pahl2001mathematical}. The definition of the level of a node becomes the maximum value of the number of direct edges between that node and the connected sinks. In the example of Figure \ref{fig:two}, nodes $4$, $5$, $8$, and $10$ are at the third level in the single flow-direction network as they can be identified as donors to donors of the sink node. However, in the case of multiple flow-direction network, only node 8 remains in that level, with nodes 5 and 10 drifting to the fourth level, and node 4 drifting to the fifth level. 

Thus, there is a many-to-many relationship among the nodes of a multiple flow-direction network compared to the one-to-many relationship in a single flow-direction network. The proposed algorithm renders the linear layout of the multiple flow-direction network considering the possibility of multiple receivers as well as donors for a node during the network traversal. This fundamental change in the node connectivity modifies the criteria of the linear ordering for an efficient implicit calculation of the erosion term in LEMs.

\section{The algorithm}
\label{S4}

The algorithm (the pseudocode is presented in \ref{appendix-sec2}) for implicitly solving the erosion term in a multiple flow-direction network starts by determining the donors of all nodes using the information about the receivers of the nodes (Section \ref{SS41}). It creates the queue (linear ordering) to process nodes, and then implicitly updates elevations for all nodes in the domain using the erosion term (Section \ref{SS42}). The proposed algorithm does not depend on the positioning of the nodes and it can be used for any regular or irregular mesh. D$\infty$ (a multiple flow-direction method) is used to compute the total drainage area ($A$) and test this algorithm.

\subsection{Determination of the donors}
\label{SS41}
The maximum number of donors possible for a node is the number of neighbors of that node. This value is eight for the rectangular grid chosen in our numerical model. We used a two-dimensional matrix $\mathcal{D}_{N\times 8}$ to store the donors' information for each node, where $N$ is the number of nodes. Information about donors of each node is obtained from the receiver array, which is assembled based on the node connectivity in a flow network. The receiver array ($\mathcal{R}_{N\times k}$) is a two-dimensional matrix where $N$ is the number of nodes and $k$ is the maximum number of receivers allowed by the flow-direction method ($k=2$ for D$\infty$). Each node is checked to assess whether it is the receiver of its neighbor. If it is, that neighbor is stored as one of the donors of the node.

\subsection{Ordering to process nodes}
\label{SS42}
The queue ($\mathcal{Q}$) is the one-dimensional data structure that contains the traversal order of nodes in the network. The sequence of nodes in $\mathcal{Q}$ is such that the elevation values of a node's receivers are already updated before that node's elevation is updated implicitly (Table \ref{table1} presents the ordering of nodes, $\mathcal{Q}$, for the multiple flow-direction network in Figure \ref{fig:two}(b)). This ordering allows computing elevation for every node in the domain implicitly with the time complexity of the algorithm varying linearly with the number of nodes. At the beginning of each time-step, nodes without receivers (sink nodes) are added to the queue and are marked as processed. We extract an element from the front of the queue and visit its donor. If all receivers of that donor are already processed, its elevation is updated implicitly by solving equation
\begin{equation}
\label{eq:erosion_term_calculation}
z_e^{p+\frac{1}{2}} = z_e^p - K a^m |\nabla z_e^{p+\frac{1}{2}}|^n \Delta t,
\end{equation}
where $z_e^p$ is elevation value of the donor at previous time-step, $z_e^{p+\frac{1}{2}}$ is the updated elevation using the erosion term, $|\nabla z_e^{p+\frac{1}{2}}|$ is the slope and $\Delta t$ is the time-step. In multiple flow-direction networks, the slope at a node can be calculated as the vector sum of two or more directions. In D$\infty$, the node being visited can have two flow-receiving neighbors \cite{tarboton1997new} - one in any cardinal direction (with $z_{c}^{p+\frac{1}{2}}$ as the updated surface elevation) and one in the adjacent diagonal direction (with $z_{d}^{p+\frac{1}{2}}$ as the updated surface elevation). If the grid spacing is $\Delta x$, the downward slope, $|\nabla z_e^{p+\frac{1}{2}}|$, is calculated as
\begin{equation}
\label{eq:slope_calc}
    | \nabla z_e^{p+\frac{1}{2}} | = \Bigg[ \Big(\frac{z_{c}^{p+\frac{1}{2}}-z_{e}^{p+\frac{1}{2}}}{\Delta x} \Big)^2 + \Big(\frac{z_{d}^{p+\frac{1}{2}}-z_{c}^{p+\frac{1}{2}}}{ \Delta x} \Big)^2\Bigg]^{\frac{1}{2}}.
\end{equation}

After obtaining the slope, the non-linear equation (\ref{eq:erosion_term_calculation}) can be solved for a node using the root-finding algorithms like Brent's method or Newton-Raphson method \cite{kiusalaas_2013, ram2009engineering}. The node is then marked as processed and is pushed into the queue. This step modifies the network traversal from the single flow-direction network since a donor of the node in the queue cannot be immediately processed until its all other receivers are processed.

\begin{table}[!hbt]
\caption{A worked example for the multiple flow-direction network shown in Figure \ref{fig:two}(b). $i$ indicates the array of nodes with matrix $\mathcal{R}$ storing receiver nodes for the respective nodes. Matrix $\mathcal{D}$ is assembled by using the information of $\mathcal{R}$ as explained in Section \ref{SS41}. $\mathcal{Q}$ displays the order of nodes for efficient implicit computation of the elevation starting from sink 1 and 6 of the network.}
\centering
\begin{tabular}{|lllllllllll|}
\hline
$i$ & 1 & 2 & 3 & 4 & 5 & 6 & 7 & 8 & 9 & 10 \\
\hline
$\mathcal{R}$ & - & 1 & 1 & 3 & 3 & - & 6 & 7 & 6 & 8 \\
              & - & - & - & 5 & 8 & - & - & - & - & 9 \\
$\mathcal{D}$ & 2 & - & 4 & - & 4 & 7 & 8 & 5 & 10 & - \\
              & 3 & - & 5 & - & - & 9 & - & 10 & - & - \\
$\mathcal{Q}$ & 1 & 6 & 2 & 3 & 7 & 9 & 8 & 10 & 5 & 4 \\
\hline
\end{tabular}
\label{table1}
\end{table}

To get the final elevation of a node, we first update the elevation by implicitly solving equation (\ref{eq:erosion_term_calculation}) using the proposed algorithm, followed by implicitly updating diffusion and uplift as
\begin{equation}
\label{eq:rest_term_calculation}
z_e^{p+1} = z_e^{p+\frac{1}{2}} + (D \nabla^2 z_e^{p+1} +U) \Delta t,
\end{equation}
where $z^{p+1}_{e}$ is the final updated elevation after a time-step. We have employed the five-point stencil second order central-difference formula for discretizing the Laplace operator ($\nabla^2$) \cite{becker2016numerical}. This results in a 5-diagonal (sparse) matrix system, which is solved using the LGMRES algorithm \cite{baker2005technique}.

\section{Numerical results}
\label{S5}

We performed numerical experiments for the square and rectangular domains with boundary nodes at fixed zero elevations. We start with a flat surface with random spatial noise as the initial topography and update elevation values over the entire domain until the topographic steady-state is reached \cite{willett2002steady}. We consider diffusion and erosion coefficients, model parameters ($m$ and $n$), and the uplift rate to be constant in space and time. In this study, we further considered the flow width is equal to the grid spacing and ignored any sub-grid resolution features. For large grid spacing, the reader is referred to the methods discussed to accurately scale the flow width present in the sub-grid scale of the discretized domain in \cite{tucker2010modelling, howard1994detachment, pelletier2010minimizing, pelletier2012fluvial}. 

\subsection{Code and solution verification}
\label{SS51}
We compared numerical solutions to the analytical solutions as a part of code verification and computed the observed level of accuracy for the solution verification \cite{oberkampf2010verification, roache1998verification, roy2005review}.

\subsubsection{Code verification with analytical solution}
\label{SS511}
For a semi-infinite domain of width $l$ with parameters $m=n=1$, the steady-state analytical solution can be obtained following \cite{bonetti2020channelization}. Assuming that the elevation decreases monotonically on the either side of divide in 1D transect and defining $x_* = x - l/2$, equation (\ref{lem_eq}) at steady-state becomes
\begin{equation}
\label{lem_eq_ss}
D\nabla^2 z+ K a \nabla z + U = 0.
\end{equation}
Equation (\ref{a_eq}) yields $a = x_*$ in 1D, which gives the final form of equation (\ref{lem_eq_ss}) as
\begin{equation}
\label{lem_eq_ss_final}
D z'' + K x_* z' + U = 0.
\end{equation}
With the boundary conditions $z'(x_*=0) = 0$ and $z(x_*=0) = z_o$ at the divide, equation (\ref{lem_eq_ss_final}) is solved as
\begin{equation}
\label{1dsolution}
z(x_*) = z_o -\frac{U x_*^2}{2 D} { }_{p}F_q \left(1,1;\frac{3}{2},2;-\frac{K x_*^2}{2 D}\right),
\end{equation}
\begin{equation}
\label{1d_dash_solution}
|z'(x_*)| = \frac{\sqrt{2U^2}}{\sqrt{DK}} Daw\Big(\frac{\sqrt{Kx_*^2}}{\sqrt{2D}}\Big),
\end{equation}
where ${ }_{p}F_q(., .; ., .; .)$ and $Daw(.)$ are the generalized hypergeometric function and Dawson function respectively \cite{abramowitz1964handbook}. Equation (\ref{1dsolution}) gives the symmetric unchannelized hillslope profile for width $l$ with divide in the middle.

\begin{figure}[!hbt]
\centering
\includegraphics[width=\linewidth]{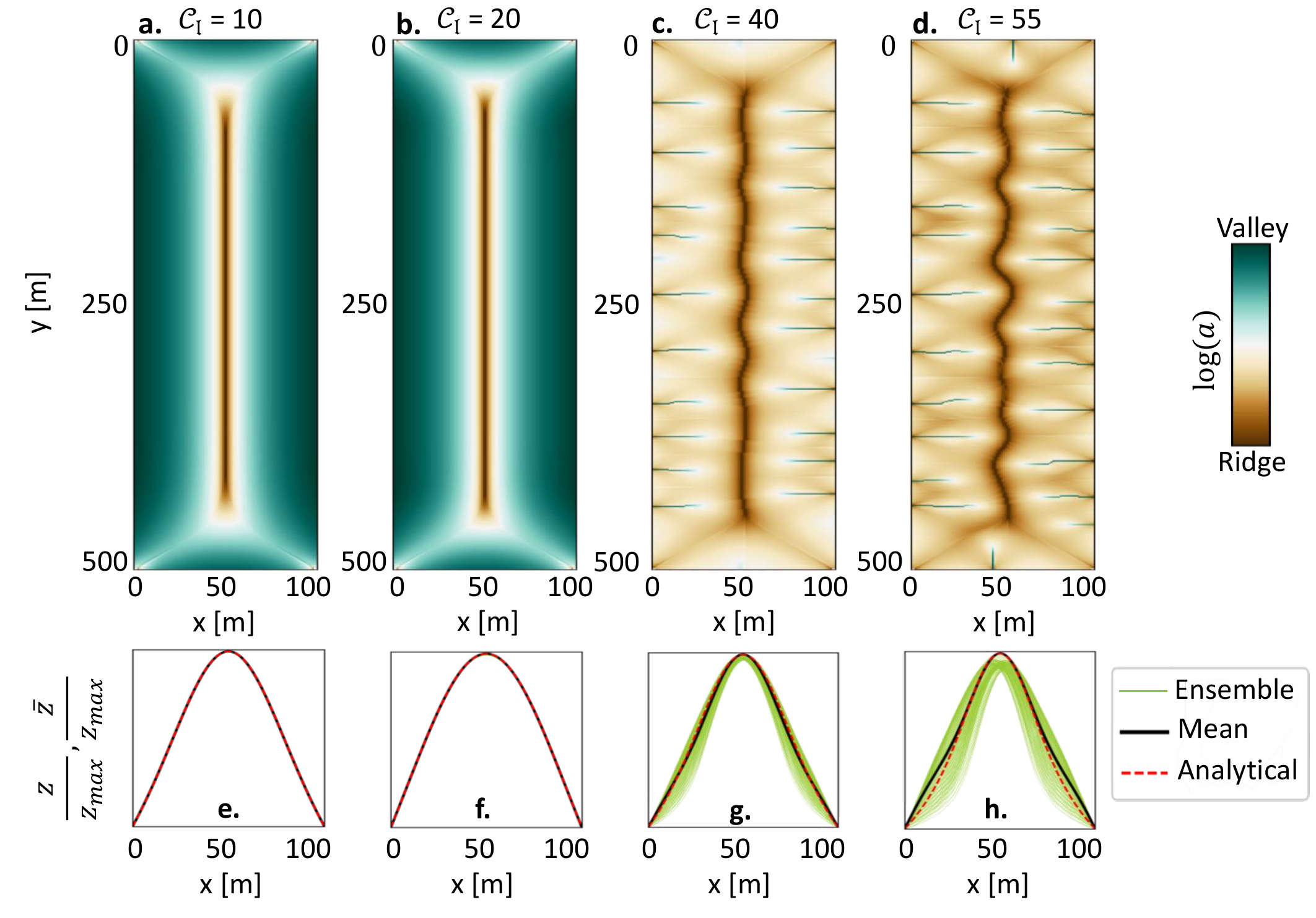}
\caption{Simulation results for the rectangular domain (width = 100 m, length = 500 m with 1 m grid spacing) and model parameters $m=n=1.0$, $D = 5.0 \times 10^{-3}$ m${}^{2}$ year${}^{-1}$, $U = 10^{-3}$ m year${}^{-1}$. (a-d): Ridge/valley network for various values of $\mathcal{C_I}$, (brown:ridge, green:valley). (a-b): before the first channel instability occurs, (c-d): after the first channel instability. (e-h): Normalized elevation profiles: Red lines are the analytical solution given by equation (\ref{1dsolution}) (representing the unchannelized case), black solid-lines are the mean elevation profile along the length and green lines show ensemble of all the profiles along the length of domain (neglecting the last 100 m of the domain on both sides).}
\label{fig:three}
\end{figure}

To compare the simulation results with the analytical solution, we considered a rectangular domain with a high aspect ratio ($l_y/l_x= 5$). A steady-state solution was obtained with the presented algorithm for $\mathcal{C_I}$ = 10, 20, 40, and 55. We compared the computed mean elevation profile along the length of the domain (neglecting the extreme sections) with the analytical solution given by equation (\ref{1dsolution}). The mean elevation profile along the length resembles the analytical profile until the first channel instability occurs (Figure \ref{fig:three}(e,f)). Only after the first channelization, the mean elevation profile starts deviating from the analytical solution as expected (Figure \ref{fig:three}(g, h)).

\subsubsection{Solution verification}
\label{SS512}
The proposed algorithm is theoretically first-order accurate in space as well as time. If high accuracy is required, a high-order scheme, such as the Crank-Nicolson temporal method, can be employed to get the second-order temporal accuracy \cite{langtangen2017finite}. The finite-difference discretization used in this study may not be suitable for the cases where sharp discontinuities (e.g., knickpoint migration) exist in the solution \cite{moin2010fundamentals, toro2009riemann}. Under such circumstances, other numerical schemes such as finite volume method may be a better choice \cite{campforts2015keeping}. In the present study, however, the governing equations contain a linear diffusion term representing soil creep. This prevents the formation of singularities in the solution and does not lead to detrimental numerical errors in the model.

\begin{figure}[!hbt]
\centering
\includegraphics[width=\linewidth]{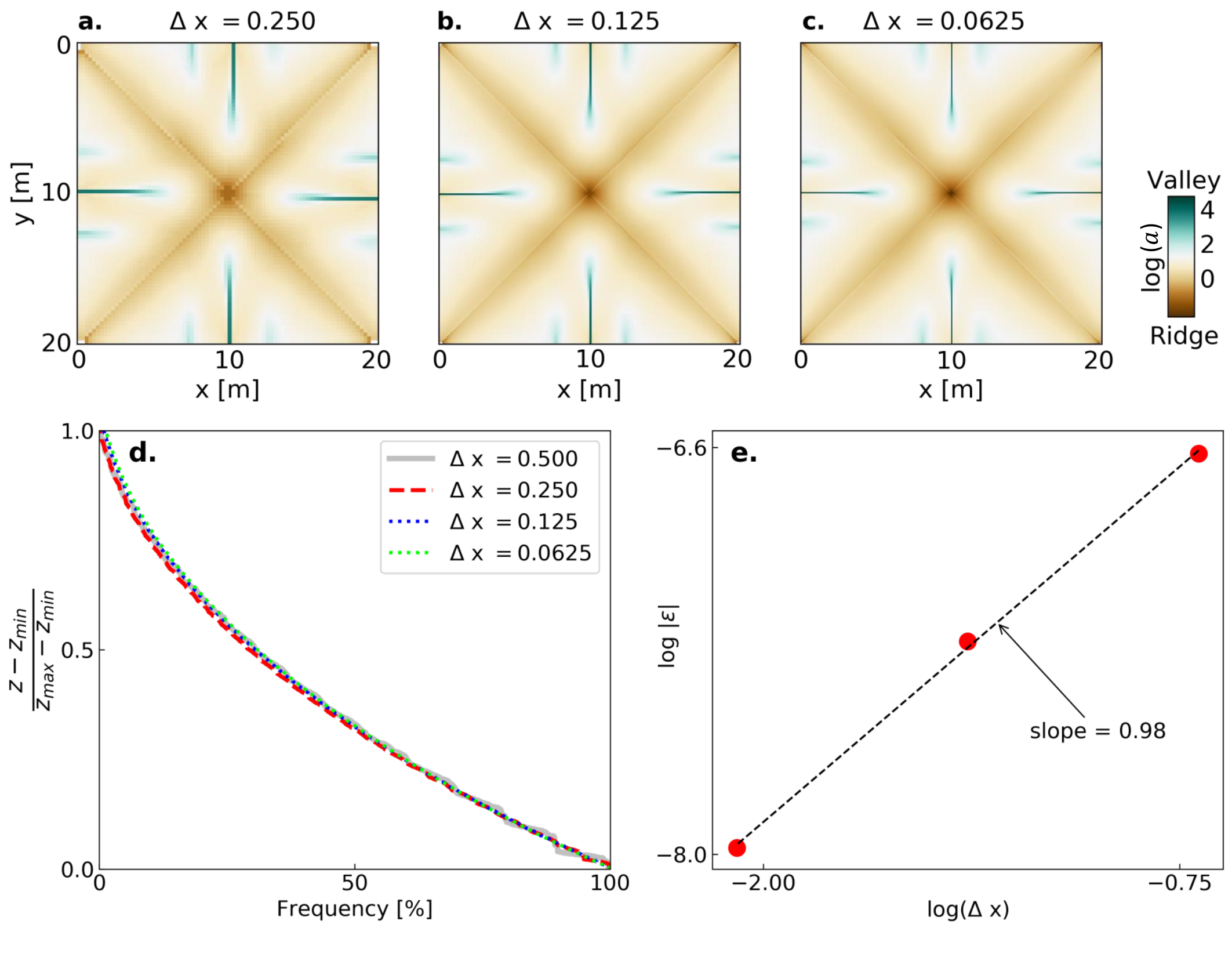}
\caption{Steady-state solutions for a square domain (side length = 20 m), $m = 0.5$, $n = 1.0$, $D = 5.0 \times 10^{-3}$ m${}^{2}$ year${}^{-1}$, $U = 5.0 \times 10^{-5}$ m year${}^{-1}$, and $\mathcal{C_I} = 62$. (a), (b) and (c) represent the ridge/valley network for $\Delta x=0.25$, $\Delta x = 0.125$ and $\Delta x = 0.0625$ respectively (brown = ridge, green = valley). (d): Normalized hypsometric curves for $\Delta x=0.5$, $\Delta x=0.25$, $\Delta x = 0.125$ and $\Delta x = 0.0625$. (e): Linear relationship between pseudo error versus grid spacing with slope equal to 0.98 indicates the first-order spatial accuracy of the implemented numerical algorithm.}
\label{fig:four}
\end{figure}

To test the accuracy of our solutions, we decreased the grid spacing (keeping model parameters and boundary conditions same) and observed the change in the numerical error as well as the spatial patterns in the steady-state landscape profiles. Solutions for two meshes $M_1$ and $M_2$, with grid spacing $\Delta x$ and $ \Delta x/2$ respectively, can be written as $f_1 = f_{exact} + \mathcal{O}(\Delta x)^p$ and $f_2 = f_{exact} + \mathcal{O}(\Delta x/2)^p$. Expanding $\mathcal{O}(\Delta x)^p$ and $\mathcal{O}(\Delta x/2)^p$ terms, and neglecting higher order terms, these equations can be written as
\begin{equation}
\label{psedo_eq_1}
f_1 = f_{exact} + C_p(\Delta x)^p, f_2 = f_{exact} + C_p(\Delta x/2)^p.
\end{equation}
Taking the difference of these two equations and taking the logarithm on both sides gives
\begin{equation}
\label{psedo_eq_2}
\log(\epsilon)= p\log(\Delta x) + C, 
\end{equation}
where $\epsilon = f_1 -f_2$ (pseudo-error) and $C$ is a constant. This means that the slope of linear plot of pseudo-error versus grid spacing gives the order of accuracy of implementation. In the spatial convergence test, we considered four grid spacing $\Delta x = 0.5$, $\Delta x/2$, $\Delta x/4$ and $\Delta x/8$ for a square domain (side length = 20 m) for $\mathcal{C_I} = 62$ (Figure \ref{fig:four}(a,b,c)). We computed pseudo-errors using mean elevation as a metric for these cases and obtained the best-fit line on a scatter plot of grid spacing vs. pseudo-error. As can been seen in Figure \ref{fig:four}(e), the slope of the best-fit line is 0.98 (close to one)  which shows the observed level of accuracy from the implementation follows the theoretical predictions. Normalized hypsometric curves for the four cases were found to be in good agreement, indicating that the proportion of land at various levels remains unaltered in the spatial convergence test (Figure \ref{fig:four}(d)).

\subsection{Single vs multiple flow-direction method and the first channelization}
\label{SS52}
\begin{figure}[!hbt]
\centering
\includegraphics[width=\linewidth]{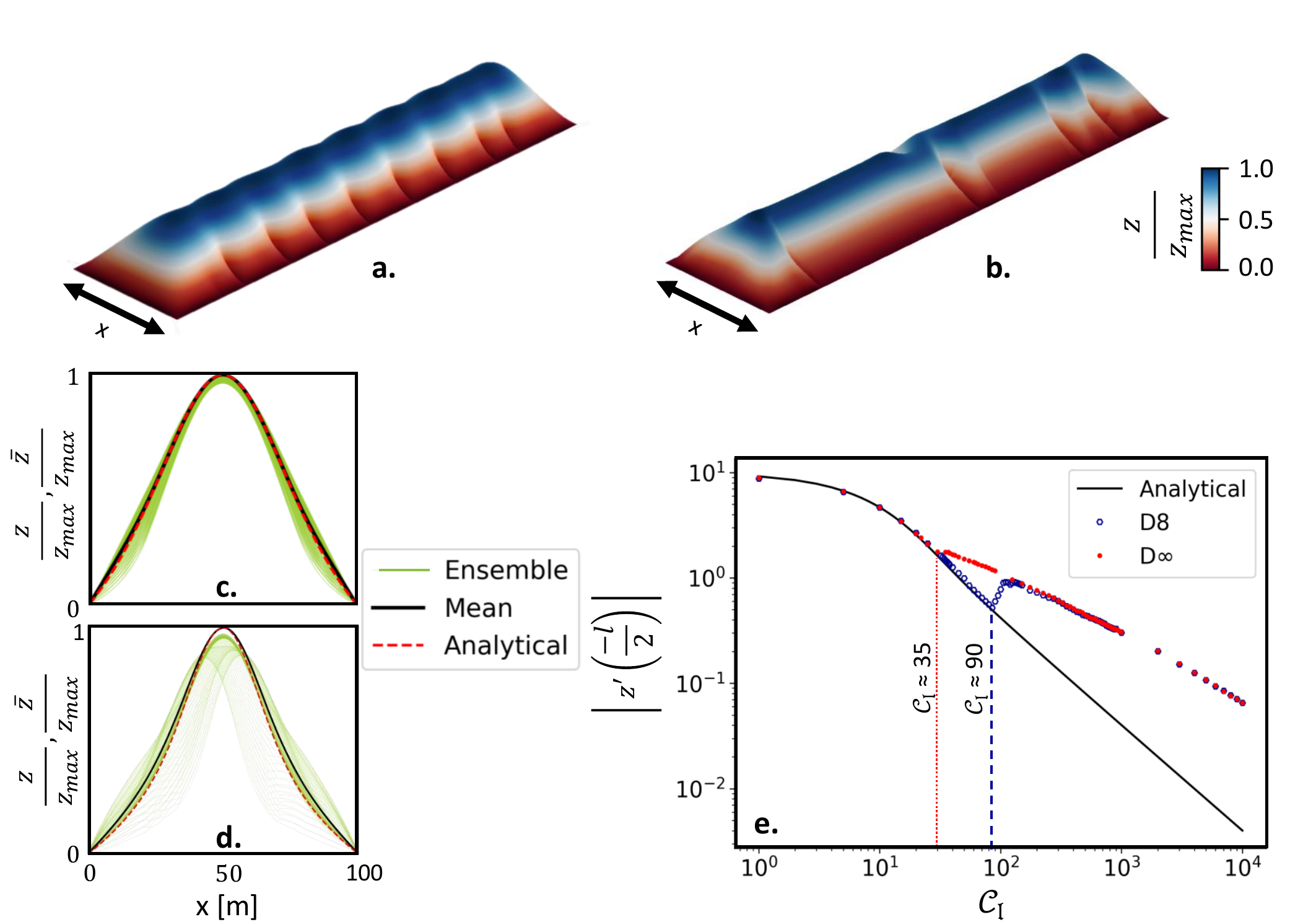}
\caption{Steady-state landscape profiles for the rectangular domain at first channel instability (width = 100 m, length = 500 m with 1 m grid spacing) with model parameters $m = n = 1.0$, $D = 5.0 \times 10^{-3}$ m${}^{2}$ year${}^{-1}$, $U = 10^{-3}$ m year${}^{-1}$. (a): $\mathcal{C_I} = 35$ for D$\infty$ method, (b): $\mathcal{C_I} = 90$ for D8 method (blue = ridge, red = valley). (c,d): Normalized elevation profiles for the landscapes shown in panels (a) and (b) respectively, red-dotted lines are the analytical solution for the unchannelized case given by equation (\ref{1dsolution}), black solid-lines are the mean elevation profile along the length and green lines show ensemble of all the profiles along the length of the domain (neglecting the last 100 m of the domain). (e): Slope at the boundary of mean elevation profile along the length for D$\infty$ (red circle), D8 (unfilled blue octagon) and the unchannelized case (black line) given by equation (\ref{1d_dash_solution}).}
\label{fig:five}
\end{figure}

Linear stability analysis on the steady-state analytical solution (equation (\ref{1dsolution})), performed by \cite{bonetti2020channelization}, shows that the first channel instability occurs for $\mathcal{C_I} \approx 37$. Here, we focused on the initiation of the first channel for a rectangular domain with a high aspect ratio ($l_y/l_x= 5$) using single and multiple flow-direction methods (D8 and D$\infty$ respectively). We also analyzed the steady-state landscape profiles for different values of $\mathcal{C_I}$ using D8 and D$\infty$ flow-direction methods. For D8, the implementation of \cite{braun2013very} algorithm in Landlab was used to compute the erosive term in the solver \cite{hobley2017creative}, while for D$\infty$, we used our proposed algorithm. In close agreement with the theoretical analysis, the first channel instability was found to occur at $\mathcal{C_I} = 35$ for D$\infty$ method, while the first channel was observed only around $\mathcal{C_I} \approx 90$ for D8 method (Figure \ref{fig:five}(a-d)).

We further compared the slope of the mean elevation profile along the length at the boundary for both the flow-direction methods with the analytical solution for the unchannelized case (equation (\ref{1d_dash_solution})). As seen in Figure \ref{fig:five}(e), the slope starts deviating for D$\infty$ when the first channel instability occurs at $\mathcal{C_I}$ around 35, while it occurs around $\mathcal{C_I} = 90$ for D8. This indicates that the transition from smooth to dissected landscape is not captured well by D8 method. Our results parallel the conclusion of the numerical investigation in \cite{gallant2011differential}, where the theoretical values of $a$ obtained from equation (\ref{a_eq}) are compared with the approximated values applying different flow-direction methods such as D8, D$\infty$ and DEMON for simple geometries. D$8$ especially gives poor results whereas D$\infty$ most accurately approximates $a$ on hillslopes \cite{gallant2011differential}. This clearly shows the inadequacy of the single flow-direction method in cases where a good approximation of the specific drainage area is needed.

\subsection{Mean elevation dynamics}
\label{SS53}
We further assessed the accuracy of the proposed algorithm by considering the transient evolution of mean elevation, for which it is possible to have an analytical expression. For a rectangular domain with fixed boundary elevations, the mean elevation is given by
\begin{equation}
\label{z_mean_eq}
\bar{z} = \frac{1}{A} \oiint z ds,
\end{equation}
where $\oiint$ represents closed surface integral and $ds$ is the infinitesimal area element in the domain having area $\mathcal{A}$.

Using equation (\ref{lem_eq}) and assuming $m=n=1$, the temporal dynamics of mean elevation can be written as 
\begin{equation}
\label{lem_mean_eq}
\frac{d \bar{z}}{d t}= \frac{1}{A} \oiint (D\nabla^2 z-K a |\nabla z| + U)ds.
\end{equation}

The divergence theorem allows us to write the first term on RHS of equation (\ref{lem_mean_eq}) as $\oint D \nabla z.\vec{n} d \Omega $, where $\vec{n}$ is the normal vector to the domain boundary ($ \Omega$). This term gives the summation of the gradient of boundary nodes along the normal to the boundaries. The second term on RHS of equation (\ref{lem_mean_eq}) can be expressed using orthogonal curvilinear coordinates ($u$, $v$) where $u$ directs along the contour lines and $v$ along the stream lines \cite{gallant2011differential, jeffreys_jeffreys_1999}. 

The length elements along $u$ and $v$ are $dw = \sqrt{L_u} du$ and  $dl = \sqrt{L_v} dv$ respectively, where $L_u = x_u^2+ y_u^2$ and $L_v = x_v^2+ y_v^2$. Further, the infinitesimal area ($ds$) in ($u$, $v$) coordinate system is  $J dv du$, where $J$ is the Jacobian defined as $|x_u y_v - x_v y_u|$. As a consequence of the orthogonality of $u$ and $v$, we have $J = \sqrt{L_uL_v}$. Using these relations, the integral equation for $a$ in ($u$, $v$) coordinate system is derived in \cite{gallant2011differential} as
\begin{equation}
\label{lem_a_int_eq}
a = \frac{1}{\sqrt{L_u}}\int\limits_{v} J dv,
\end{equation}
and the slope is defined as
\begin{equation}
\label{lem_gradz_int_eq}
|\nabla z| = -\frac{1}{\sqrt{L_v}}\frac{\partial z}{\partial v}.
\end{equation}
Substituting these expressions in the second term on RHS of equation (\ref{lem_mean_eq}), we get
\begin{equation}
\label{int_flov_1}
\frac{-K}{A} \oiint a |\nabla z| ds = \frac{K}{A} \int\limits_{u}\int\limits_{v} \left(\frac{1}{\sqrt{L_u}} \int\limits_{v} J dv'\right)\frac{1}{\sqrt{L_v}} \frac{\partial z}{\partial v} J dv\ du.
\end{equation}
Using integration by parts, equation (\ref{int_flov_1}) is further modified as
\begin{equation}
\label{int_flov_2}
\frac{-K}{A} \oiint a |\nabla z| ds = \frac{K}{A} \int\limits_u \left[  z  \left( \int\limits_{v} J dv'\right) \Bigg |_{v_0}^{v_{b}} - \int\limits_v \frac{\partial}{\partial v}\left (  \int\limits_{v} J dv' \right) z dv \right] du.  
\end{equation}
where $v_0$ and $v_b$ are the along-stream coordinates at the initiation and end of each stream line. At the initiation points ($v=v_0$), we have $a=0$ and thus $\int\limits_{v} J dv'=0$. At the end points ($v=v_b$), we have imposed the boundary condition of $z=0$. Given these conditions, the first term in the integrand on RHS of equation (\ref{int_flov_2}) is zero for any stream line $u$ ($z  \left( \int\limits_{v} J dv'\right) \Bigg |_{v_0}^{v_{b}}=0$), which simplifies the equation as
\begin{equation}
\label{int_flov_3}
\frac{-K}{A} \oiint a |\nabla z| ds = -K  \int\limits_u\int\limits_v z J dv du = - K \bar{z}. 
\end{equation}

\begin{figure}[!hbt]
\centering
\includegraphics[width=\linewidth]{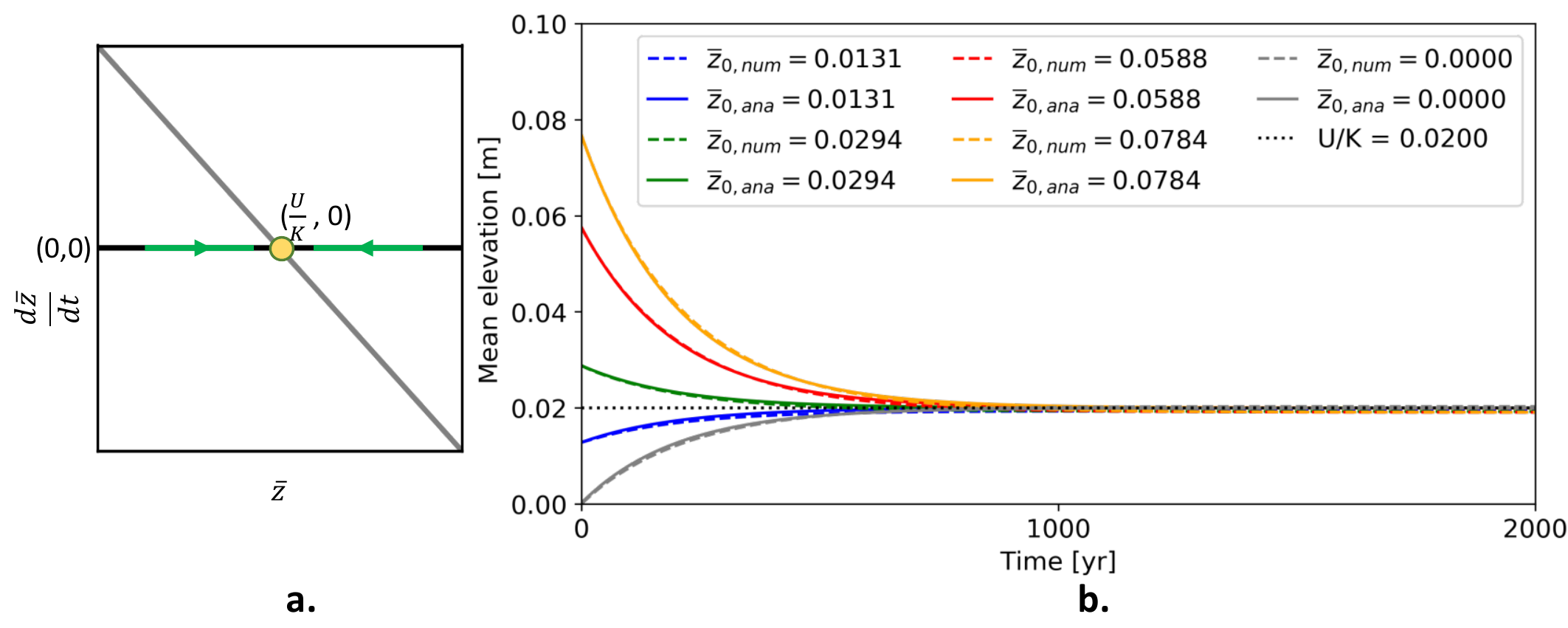}
\caption{Mean elevation dynamics. (a): The flow (green arrows) for the dynamical system given by equation (\ref{mean_ele_f_eq}) with the stable fixed point (yellow). (b): The analytical solution (solid lines) given by equation (\ref{mean_ele_f_eq_time}) compared with the simulation results (dashed lines) using our algorithm for D$\infty$ method in a square domain of side length 100 m with 1 m grid spacing at $\mathcal{C_I} = 10^6$, $m=n=1$, $D = 5.0 \times 10^{-5}$ m${}^{2}$ year${}^{-1}$, $U = 10^{-4}$ m year${}^{-1}$ and $U/K = 0.02$ starting from different values of $\bar{z}_o$.}
\label{fig:six}
\end{figure}

For a high value of $\mathcal{C_I}$ ($D \rightarrow 0$), the contribution from the first term on RHS of equation (\ref{lem_mean_eq}) is negligible as the slope at boundary decreases on increasing the value of $\mathcal{C_I}$ (Figure \ref{fig:five}(e)), which makes the temporal variation of mean elevation as
\begin{equation}
\label{mean_ele_f_eq}
\frac{d \bar{z}}{d t} \approx -K \bar{z} + U.
\end{equation}
The solution is
\begin{equation}
\label{mean_ele_f_eq_time}
|U-K \bar{z}(t)|= |U-K \bar{z}_o| e^{-Kt},
\end{equation}
where $\bar{z}_o$ is the mean elevation at time $t=0$, $\bar{z}(t) $ is the value at any time $t$. Thus, the landscape reaches steady state with mean elevation value reaching $U/K$ (Figure \ref{fig:six}(a)). 

We considered a square domain (side length = 100 m, $U/K = 0.02$, $\mathcal{C_I} = 10^6$) to compare the temporal variation of mean elevation obtained from the numerical solutions with the derived analytical expression. We simulated the numerical model starting from different initial values of mean elevation ($\bar{z}_o$) and observed the temporal trajectory. As shown in Figure \ref{fig:six}(b), the steady-state mean elevation for the numerical algorithm reaches the value of $U/K$ for any $\bar{z}_o$. There is also a good match between the numerical and analytical trajectories of mean elevation, indicating accurate transient solutions provided by the presented numerical algorithm using D$\infty$ method.

\subsection{High values of \texorpdfstring{$\mathcal{C_I}$}{Ci}}
\label{SS54}
\begin{figure}[!hbt]
\centering
\includegraphics[width=\linewidth]{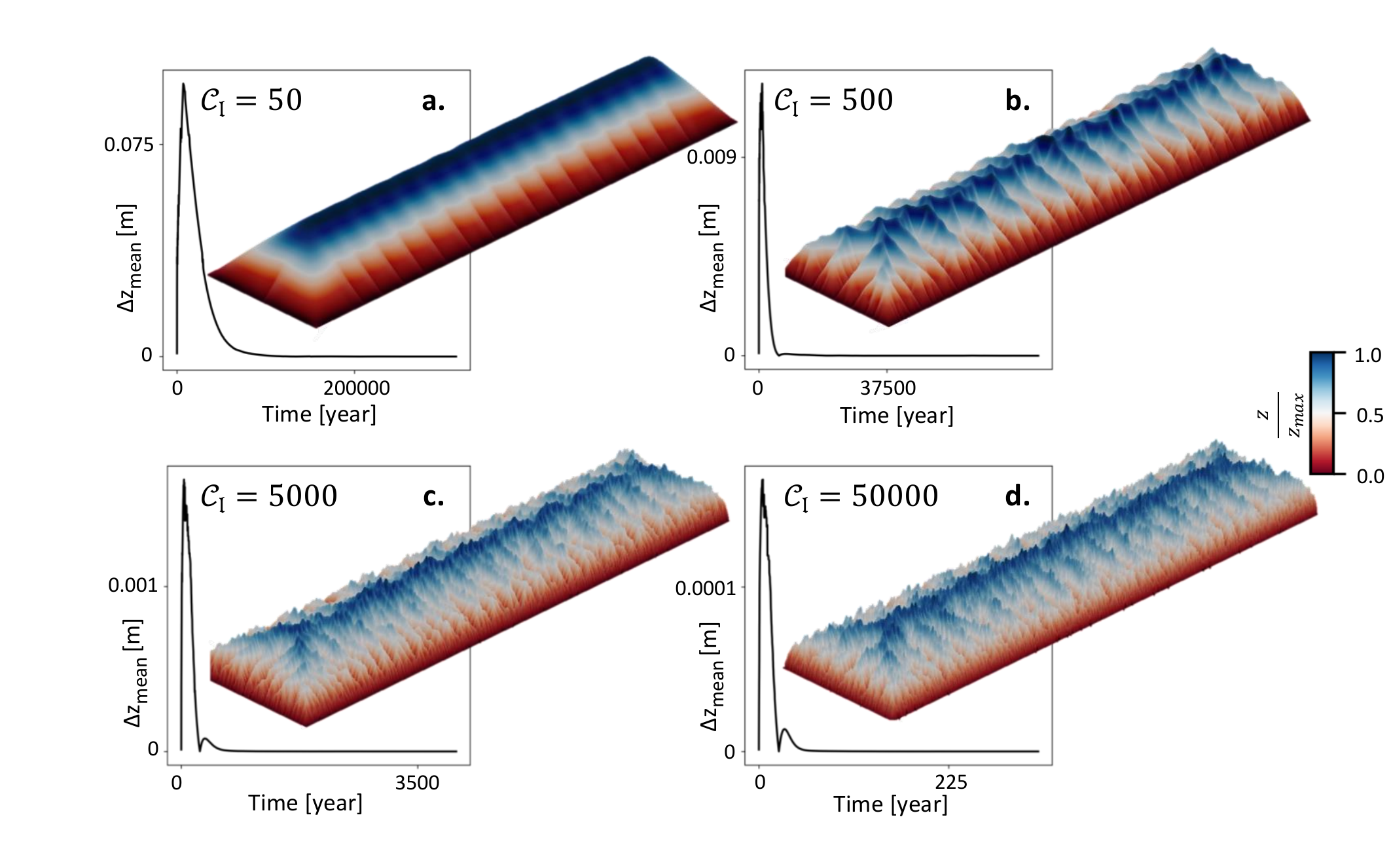}
\caption{(a,d): Steady-state landscape profiles for the rectangular domain (width = 100 m, length = 500 m with 1 m grid spacing) with model parameters $m = 0.5$, $n = 1.0$, $D = 5.0 \times 10^{-3}$ m${}^{2}$ year${}^{-1}$, $U = 10^{-3}$ m year${}^{-1}$ for various values of $\mathcal{C_I}$ (blue = ridge, red = valley). Insets represent the variation of change in mean elevation value over the time during the simulations. The average $\Delta t$ ranges from 120 years for small $\mathcal{C_I}(=50)$ to 0.1 years for high $\mathcal{C_I} (= 50000)$. The number of time-steps remain around 3000-4000 for various $\mathcal{C_I}$ values as the high $\mathcal{C_I}$ case reaches steady-state particularly faster than small $\mathcal{C_I}$ case.}
\label{fig:seven}
\end{figure}

A major issue with the explicit solvers using a multiple flow-direction method is the limitation on the time-step size for the erosion term as per the stability criteria. This limitation poses a practicality constraint on obtaining numerical solutions for the high values of $\mathcal{C_I}$. Our algorithm resolves this issue by using an efficient implicit computation of the erosion term, which does not impose any restrictions on the maximal time-step value for this part of the model. To illustrate this point, the solver was employed to get steady-state landscape profiles for $\mathcal{C_I} = 50$ to $\mathcal{C_I} = 50,000$. Figure \ref{fig:seven} represents the steady-state solutions for the rectangular domain ($l_y/l_x= 5$), along with the variation of change in mean elevation value over consecutive time-steps (shown in respective insets) for different values of $\mathcal{C_I}$. Increasingly complex channel forms are obtained for high values of $\mathcal{C_I}$ \cite{bonetti2020channelization}. After the initial period of channel initiation, change in the mean elevation decreases smoothly until topographic steady-state is reached, indicating the absence of any numerical instability engendered by the proposed algorithm. These results demonstrate high efficiency and robustness of the solver for a varied range of parameter values.

\section{Conclusion and discussion}
\label{S6}

Extending the interesting contributions of \cite{barnes2019accelerating, braun2013very}, we proposed an efficient algorithm for the multiple flow-direction network to compute implicitly the erosion term of equation (\ref{lem_eq}) in the numerical model of the detachment-limited landscape evolution dynamics. The algorithm depends only on the connectivity among the nodes in the multiple flow-direction network rather than their spatial positions, which makes it adaptable to any irregular mesh. The lack of constraint on the time-step for updating the elevation by the erosion term offers a way to obtain accurate steady-state solutions for the wide range of $\mathcal{C_I}$. In particular, the numerical solutions obtained closely follow the theoretical predictions of channel instability when approximating $a$ using D$\infty$ method. The mean elevation dynamics obtained by the numerical solution is also in good agreement with the theoretical analysis.

Instead of using flow-direction methods to compute the specific drainage area, recent contributions directly solve specific drainage-area equation (\ref{a_eq}) for a Digital Elevation Model \cite{qin2017efficient}. LEM however requires calculating the specific drainage area at every time-step, making the approach computationally expensive. The water-flow equation can also be made more detailed than our minimalist model, including, for example, water diffusion over the landscape \cite{fowler2011mathematical}. Multiple flow-direction methods, such as M8 method, have been proposed for such cases \cite{qin2007adaptive, quinn1991prediction}. 

We are currently working to link our results with those of optimal channel networks \cite{banavar2001scaling, rinaldo2014evolution} as well as optimal transport problems \cite{bergamaschi2019spectral} and extend the algorithm to efficiently simulate the vascularization and branching problems in 3D domains. 

\section{Acknowledgements}
\label{S7}
The authors acknowledge support from the US National Science Foundation (NSF) grants EAR-1331846 and EAR-1338694, and BP through the Carbon Mitigation Initiative (CMI) at Princeton University. A.P. and M.H. also acknowledge the support from the Princeton Institute for International and Regional Studies (PIIRS) and the Princeton Environmental Institute (PEI).

The authors are pleased to acknowledge that the simulations presented in this article were performed on computational resources managed and supported by Princeton Research Computing, a consortium of groups including the Princeton Institute for Computational Science and Engineering (PICSciE) and the Office of Information Technology's High Performance Computing Center and Visualization Laboratory at Princeton University.

Well-commented source code and the simulation results discussed in the paper are available at \url{https://github.com/ShashankAnand1996/LEM}.

\bibliographystyle{elsarticle-num}

\bibliography{reference}
\appendix
\section{M8 flow-direction method}
\label{appendix-sec1}
In the mathematical formulation of LEM, we assumed that the water goes in the direction of steepest descent as indicated by equation (\ref{a_eq}). D8 (single) and D$\infty$ (multiple) flow-direction methods follow the same concept with D$\infty$ splitting the flow between neighboring receivers only when the flow direction does not coincide with directions pointing toward neighbors (cardinal and diagonal directions in a rectangular grid). We therefore kept the discussion in the main text up to these methods to compare numerical results with analytical predictions using governing equations (\ref{lem_eq}) and (\ref{a_eq}). 

\begin{figure}[!hbt]
\centering
\includegraphics[width=\linewidth]{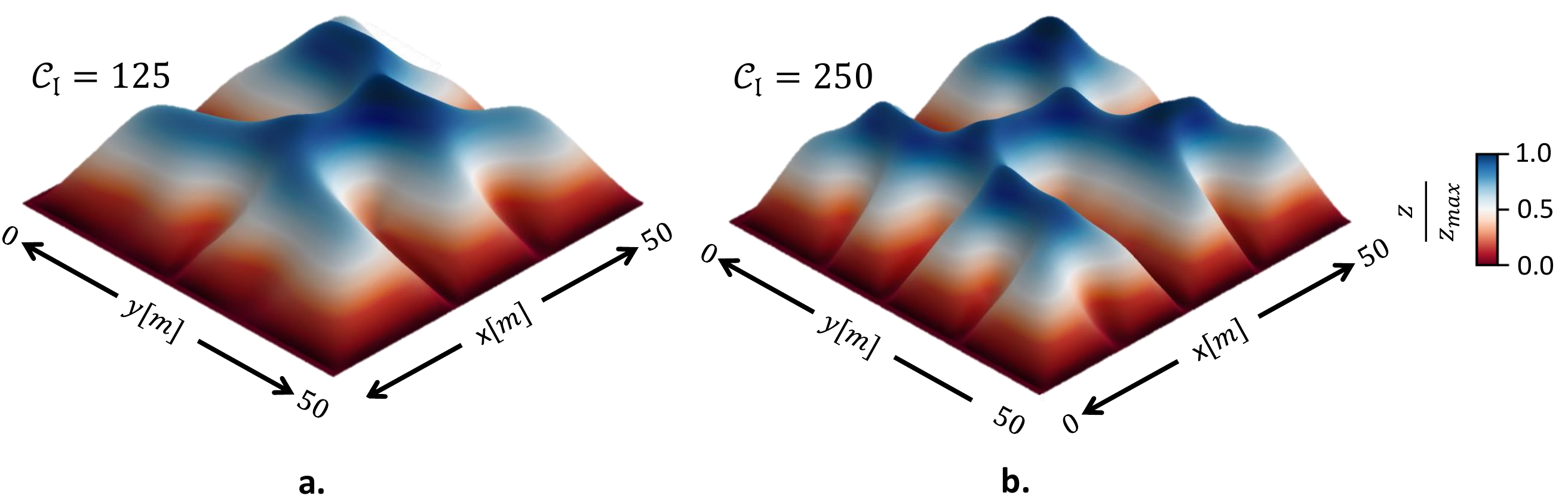}
\caption{Simulation results using M8 flow-direction method for computing $a$ in a square domain of side length 50 m with 1 m grid spacing and parameters $m = n = 1.0$, $D = 5.0 \times 10^{-5}$ m${}^{2}$ year${}^{-1}$, $U = 1.0 \times 10^{-3}$ m year${}^{-1}$ (brown = ridge, blue = valley). (a): $\mathcal{C_I} = 125$ with average $\Delta t = 8896$ years until steady state is reached at $6.95\times 10^{6}$ years. (b): $\mathcal{C_I} = 250$ with average $\Delta t = 3334$ years until steady state is reached at $4.53\times 10^{6}$ years.}
\label{fig:eight}
\end{figure}

Other multiple flow-direction methods, such as M8 method, distribute the flow to all downstream neighbors from a node where the proportion of flow is decided based on various matrices (slope proportion, some power of slope proportion etc.) \cite{qin2007adaptive, quinn1991prediction}. From the numerical point of view, the presented algorithm can be applied to any multiple flow-direction method. To show the scope of the presented study, we applied the algorithm using M8 method to obtain steady-state solutions for a square domain (Figure \ref{fig:eight}). The flow proportion received by the downstream neighbors of a node was decided based on slope proportion in the Landlab environment \cite{hobley2017creative}. As shown in Figure \ref{fig:eight}, the increasing value of $\mathcal{C_I}$ (from 125 to 250) amplifies the channelization in the square domain as expected.

\section{Pseudocode}
\label{appendix-sec2}
The details about the implementation of the proposed algorithm are presented in this appendix. The algorithm is written for serial programming as a Python function that is compatible with the modeling environment provided by Landlab \cite{hobley2017creative}. By keeping track of nodes at the same level, it can be easily extended for parallel programming. Algorithm (\ref{al1}) generates the queue ($\mathcal{Q}$) for the input flow network and employs algorithm (\ref{al2}) to update elevation for any node implicitly using the erosion term.

$\mathcal{D}_{N\times 8}$ is a two-dimensional array that is used to store the donors' information for each node, where $N$ is the number of nodes. Another way of doing this is using the adjacency list which has the space complexity of $\mathcal{O}(V + E)$, where $V$ is the number of vertices (nodes) and $E$ is the total number of edges in the network \cite{cormen2009introduction}. This approach is extremely useful when the number of neighbors is large and the graph is sparse. $S$ is the flag used in receiver array ($\mathcal{R}$) to indicate sink nodes at each time-step. A one-dimensional array ($\mathcal{Z}_{N \times 1}$) is employed to store elevation values for all the nodes in row-major order, i.e., if the x-coordinate of the node is $i$ and y-coordinate is $j$, its location in $\mathcal{Z}$ is $N_c \times j + i$ ($N_c$ is length and $N_r$ is width of the rectangular grid). $\mathcal{A}_{N \times 1}$ is the one-dimensional array that stores the specific drainage area in row-major order for each node in the domain.

 In the queue, a new element is inserted at the end ($\mathcal{Q}.push$), and the first element is deleted ($\mathcal{Q}.pop$) following first-in first-out order (FIFO) \cite{cormen2009introduction}. We use $\mathcal{P}_{N \times 1}$, a Boolean array, to mark if the node has been processed or not (one indicating the processed node). This array helps to identify if the elevation of a node being visited can be updated implicitly or not. This step is necessary as nodes in a multiple flow-direction network have multiple receivers.

\label{Algorithms}
\begin{algorithm}[b!]
\caption{Generate the queue}
\label{al1}
\begin{algorithmic} 
\FOR {$i \in [1,N] $}
     \IF{$\mathcal{R}[i] =$ S}
         \STATE $\mathcal{Q}.push(i)$
         \STATE $\mathcal{P}[i] \gets 1$
     \ENDIF
 \ENDFOR
 \WHILE{$\mathcal{Q}.size() > 0$}
 \STATE  $n \gets \mathcal{Q}.pop(1)$
 \FOR{donor $d_o$ in $\mathcal{D}[n]$}
 \IF{$\mathcal{P}[d_o] = 0$ \AND $\mathcal{P}[\mathcal{R}[d_0,r_1]] = 1$ \AND $\mathcal{P}[\mathcal{R}[d_0,r_2]] = 1$}
 \FOR {$r_i \in [1,2] $}
     \IF{$\mathcal{R}[d_o,r_i]$ \% $N_c == d_o$ $\%$ $N_c$ \OR $\mathcal{R}[d_o,r_i] / N_c  == d_0/N_c$}
         \STATE $z_c \gets \mathcal{Z}[r_i]$ 
     \ELSE
         \STATE $z_d \gets \mathcal{Z}[r_i]$
     \ENDIF
 \ENDFOR
 \STATE call function $slope(d_o,z_c,z_d)$
 \STATE $\mathcal{Q}.push(d_o)$
 \ENDIF
\ENDFOR
\ENDWHILE
\end{algorithmic}
\end{algorithm}

\begin{algorithm}[]
\caption{Calculation of the slope using equation (\ref{eq:erosion_term_calculation})}
\label{al2}
\begin{algorithmic} 
\STATE \textbf{function} $slope(d_o,z_c,z_d)$:
     \STATE $z_0 \gets \mathcal{Z}[d_o]$
     \IF{$z_c$ and $z_d$ not None}
         \STATE $z_e \gets z_o - K. (\mathcal{A}[d_o])^m( (z_e - z_c)^2 + (z_{c} - z_{d})^2 )^{(n/2)}. \Delta t / \Delta x^n$

     \ELSIF{$z_c$ not None}
        \STATE $z_e\gets z_o - K. (\mathcal{A}[d_o])^m .(z_e - z_c)^n . \Delta t / \Delta x^n$
     \ELSE
         \STATE  $z_e \gets z_o - K. (\mathcal{A}[d_o])^m .(z_e - z_d)^n . \Delta t / (\sqrt{2}*\Delta x)^n$
     \ENDIF
     \STATE $\mathcal{Z}[d_o] \gets z_e$
     \STATE \textbf{end function}
\end{algorithmic}
\end{algorithm}
\end{document}